\documentclass[10pt,journal,compsoc]{IEEEtran}

\usepackage[nocompress]{cite}
\usepackage[pdftex]{graphicx}
\DeclareGraphicsExtensions{.pdf,.png,.jpg,jpeg}
\usepackage[normalem]{ulem}
\usepackage{color}
\usepackage{url}
\usepackage{xspace}
\usepackage{wrapfig}
\usepackage{amsmath}
\usepackage{listings}

\newcommand{\Paco}{{\sc \textbf{Paco}}\xspace}

\begin{document}

\title{{\sc Paco}: A System-Level Abstraction for On-Loading Contextual Data to Mobile Devices}

\author{Nathaniel~Wendt,~\IEEEmembership{Student Member,~IEEE,}
        Christine~Julien,~\IEEEmembership{Senior Member,~IEEE}
\IEEEcompsocitemizethanks{\IEEEcompsocthanksitem N. Wendt and C. Julien are with the Department of Electrical and Computer Engineering, The University of Texas at Austin.\protect\\
E-mail: \{nathanielwendt, c.julien\}@utexas.edu}
}

\IEEEtitleabstractindextext{%
\begin{abstract}
Spatiotemporal context is crucial in modern mobile applications that utilize increasing amounts of context to better predict events and user behaviors, requiring rich records of users' or devices' spatiotemporal histories. Maintaining these rich histories requires frequent sampling and indexed storage of spatiotemporal data that pushes the limits of resource-constrained mobile devices. Today's apps offload processing and storing contextual information, but this increases response time, often relies on the user's data connection, and runs the very real risk of revealing sensitive information. In this paper we motivate the feasibility of on-loading large amounts of context and introduce \Paco (Programming Abstraction for Contextual On-loading), an architecture for on-loading data that optimizes for location and time while allowing flexibility in storing additional context. The \Paco API's innovations enable on-loading very dense traces of information, even given devices' resource constraints. Using real-world traces and our implementation for Android, we demonstrate that \Paco can support expressive application queries entirely on-device. Our quantitative evaluation assesses \Paco's energy consumption, execution time, and spatiotemporal query accuracy. Further, \Paco facilitates unified contextual reasoning across multiple applications and also supports user-controlled release of contextual data to other devices or the cloud; we demonstrate these assets through a proof-of-concept case study. 
\end{abstract}
}

\maketitle

\IEEEraisesectionheading{\section{Introduction}\label{sec:introduction}}

\IEEEPARstart{S}{patiotemporal} data is increasingly essential to mobile applications, both those that we already use on a daily basis and those envisioned as part of a future Internet of Things. These applications rely on understanding the relationships between users and their surroundings and the ways in which those relationships change over space and time. Essential to mapping this relationship is a basic understanding of the user's spatiotemporal history---basically a timestamped trace of the user's device through space.

Implementing applications that use such rich spatiotemporal context information is possible given today's technical capabilities~\cite{michel16:from}. Today's mobile devices already largely collect this information, whether by applications that collect spatiotemporal traces explicitly, such as fitness and lifestyle apps\footnote{\url{http://www.mapmyrun.com}, \url{https://www.moves-app.com}}, by applications that collect (potentially spotty) spatiotemporal traces implicitly, for example via social network checkins, or, more insidiously, by service providers themselves. Furthermore, it has been shown that when users are presented with the frequency that applications request contextual information and sensor data, they are likely to adjust or revoke the permissions~\cite{almuhimedi2015your}.  However, in most of these examples, the user's spatiotemporal information is {\em off-loaded} to the cloud. This not only introduces communication and resource (e.g., energy) costs because it requires maintaining a persistent connection to the Internet, but, more importantly, it releases highly sensitive information to the control of a third party who may not be completely trustworthy or who  may be compelled to share the data. This storage pattern can also artificially limit the granularity of data collected based on a desire to limit the overhead of offloading the data. Further, these approaches are ``siloed'' such that each consumer of spatiotemporal data collects its own traces, meaning that many efforts may be redundant.

The few approaches that do not offload all of this data are not intelligent in their ability to reduce the storage footprint~\cite{kjaergaard11:energy} (e.g., they simply statically decrease sampling rates, or, worse, completely disable location services) and they target individual applications in a siloed fashion. Other approaches have created special purpose operating systems and devices that silo private information entirely\footnote{\url{https://www.silentcircle.com/products-and-solutions/devices/}}.

We instead propose a fundamental shift in how spatiotemporal-aware mobile applications are developed, by {\em on-loading} storing and querying contextual data, still making expressive spatiotemporal data available to applications but in a user-controlled way on-device. That is, we propose that responsibility and ownership of intensely personal spatiotemporal data be brought back {\em onto} the user's device and under the explicit and individual control of the user. This requires making it possible to store and query a device's spatiotemporal data using only the device's resources without any external infrastructure or communication. This on-loading is feasible given today's mobile device capabilities and will only become more so as capabilities improve.   The fundamental principle underlying our approach is that contextual data should exist in a raw form only on the device where it is collected and that decisions about how and when to share detailed contextual data should be retained by the owner of the data.  Our approach, termed \Paco, for Programming Abstraction for Contextual Onloading, makes it possible for an off-the-shelf commodity mobile device to maintain a compact yet comprehensive store of its own (spatiotemporal) contextual history, directly accessible to applications on the device. Note that while we borrow concepts from spatial and moving object databases, \Paco only stores the history of a single user.  In our vision of a future, \Paco is integrated with the OS as a system service and provides applications running on the OS the {\em only} available window into location services. Specifically, in this vision, applications would no longer have unfettered access to the device's precise location information; instead all information about a device's spatiotemporal trajectories would be available only through the \Paco interfaces. Because of this, it is important that \Paco provide a sufficiently expressive API to not over-limit application capabilities.

In addition to providing direct user control, this on-loaded contextual data store also enables unifying contextual data across applications.  Rather than maintaining multiple application-specific contextual ``silos'' (as is the case today, although typically using the cloud), we create a single contextual data abstraction that can be shared across applications.

Supporting individual users and devices is just one potential use of rich contextual data. Applications may also benefit tremendously from sharing snapshots of contextual data with other nearby devices, either via direct (device-to-device) sharing or through hyper-localized cloudlet support~\cite{satyanarayanan09:cloudlets}. While the humans about whom this data is generated are unlikely to be willing to share the raw data publicly, they are often willing to share with other nearby, even unknown, users~\cite{jones08:geographic}. Using basically the same abstractions that \Paco uses to reduce the storage footprint, \Paco also creates expressive options for sharing a spatiotemporal history that releases varying amounts of personal information. Our approach supports sharing contextual data, allowing users to retain control over how, when, and what is shared.

Of course, while on-loading spatiotemporal data storage and querying has many benefits, there are also potential drawbacks. The storage space and energy consumption on the mobile device are very important since we target users' mobile phones, which must support many varied daily tasks. Further, \Paco achieves on-loading of spatiotemporal data by intelligently determining what data is most important to maintain; because of the algorithms involved, the user must explicitly trade-off the fidelity or granularity of the stored data for the costs incurred in storing it. Therefore, as part of our evaluation of \Paco, we study both the costs and quality of the resulting on-loaded data store and their potential impacts on applications.

{\bf Contribution.} In summary, this paper's primary contribution is the introduction of the \Paco abstractions and associated system service implementation. At a basic level, \Paco completely onloads a user's spatiotemporal data storage to the user's personal device. As a result of onloading, concerns of mobile device resource constraints must be taken into account. To make this tractable, \Paco uses the semantics of the spatiotemporal data itself, along with high-level guidance from the user, to determine what to store in order to maintain sufficient spatiotemporal fidelity. This brings two primary benefits. First, the user retains individual, direct, physical control over the stored spatiotemporal information---no third party service ``owns'' or ``controls'' this highly personal information. Second, the user's own applications on the personal device can share the same spatiotemporal context, breaking down the walls of traditional on-device application silos. \Paco relies on existing structures (i.e., the R-Tree and the k-d tree, described in Section~\ref{sec:relatedwork}) and uses them in ways such that their behavior and interaction is optimized for both energy and storage concerns when used to capture highly detailed spatiotemporal information on a mobile device.

To demonstrate how \Paco can act as an abstracted location services to better protect the user, we present \Paco 
\ {\em access profiles}, which enable users to specify how and when to share filtered, fuzzed, or otherwise lossy representations of the user's own spatiotemporal context information. For each application, the user can specify an access profile controlling the query parameters, effectively controlling the lossiness of the released data.

The remainder of this paper is organized as follows. We begin in the next section by providing a high-level system model that overviews the concrete contributions embedded in \Paco. Section~\ref{sec:relatedwork} uses this model to state \Paco's contributions in the context of related efforts, explaining why this existing work fails to achieve the \Paco vision. We also highlight some key pieces of background information necessary to understand \Paco's abstractions. Section~\ref{sec:structure} presents the details of \Paco and the \Paco API. We present evaluations of \Paco in Sections~\ref{sec:benchmarking} and~\ref{sec:casestudy}; the former provides quantitative assessments of \Paco, while the latter demonstrates \Paco's application-level capabilities through use cases. Section~\ref{sec:discussion} discusses extension of \Paco and addresses some of its limitations, and Section~\ref{sec:conclusions} concludes.

\section{A Motivating System Model}\label{sec:systemmodel}
This paper describes the \Paco API and its supporting infrastructure, which together enable on-loading spatiotemporal data and rich application-level queries over that data. \Paco fits in a larger system context that makes contextual data accessible not only to local applications but also to nearby devices, opportunistically connected via device-to-device communication, and to the larger world through cloud off-loading. While this paper does not focus primarily on the elements of the \Paco API used for this sharing of the on-loaded spatiotemporal information, the broader system model includes these capabilities.

\begin{figure}[!h]
	\centering
	\includegraphics[width=.75\columnwidth]{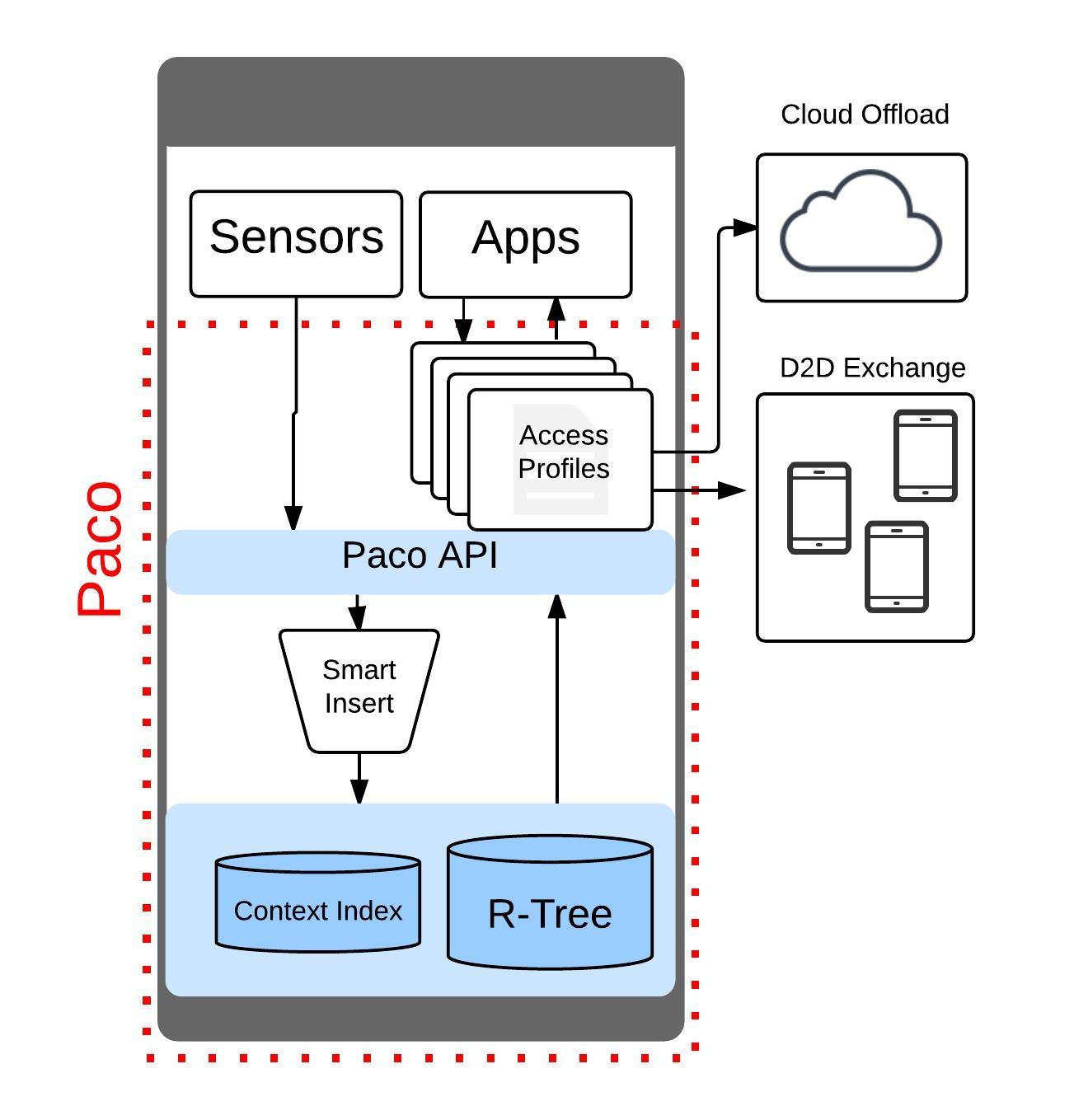}
	\caption{ System architecture; dotted region is {\em Paco}'s contributions}
	\label{fig:system-architecture}
\end{figure}

Fig.~\ref{fig:system-architecture} shows \Paco in this larger context. At the base layer of this architecture, we rely on existing implementations that can support a spatiotemporal database (``Internal Structure'' in the figure, which we implement using an {\sf R-Tree}~\cite{guttman1984r}) that can store frequent updates of time- and location-stamped observations of context. Elements are indexed within the internal structure based on their location coordinates and time stamps; each element in the internal structure is also mapped to one or more elements in the {\sf Context Index}, indicating the semantic information associated with the referenced space and time. For example, an observation might be described as: ``at 5:10pm the user is running 2 miles starting from his house (actual gps coordinates)'' or ``at 6:00am the user purchased a coffee at the cafe and gave the cafe a 5 star rating''.  For each data item, the spatiotemporal data forms the base to which additional context (e.g., that the current activity is running or that the user purchased coffee or gave a particular rating) can be added. Observations are captured by applications in user-space (e.g., via the {\sf Apps} and {\sf Sensors} components in Fig.~\ref{fig:system-architecture}).  Capturing context may require explicit user interaction, e.g., a user inputs a review of a cafe, while other context can be automatically captured in the background through sensor activity, e.g., the ambient nose level at a cafe.

The \Paco API provides access to the stored data in a manner tailored to mobile applications. At the most basic level, the user's personal applications on the device can freely query the stored spatiotemporal data, retrieving any view of the raw data desired.  Practically, users will likely want to exercise control over release of their raw contextual data even to on-device apps as well as for device-to-device (D2D) exchange and cloud offloading. For this reason, \Paco uses {\em access profiles} to further filter or otherwise alter or constrain the contextual information released from \Paco. These access profiles are user- or device-defined and can take into account the user's current situation (e.g., physical environment), his social network, etc. These access profiles can govern the granularity of the information released (e.g., releasing only aggregate ``coverage'' views of large areas) or the spaces and times about which information is released (e.g., releasing fairly detailed information about the user to his colleagues in his work environment or releasing any information that is more than a week old). These access profiles can be applied whether the data is being shared in the local environment (e.g., through D2D exchanges) or to offload contextual data to the cloud, for example offloading sufficiently dated information or lossy summaries that obfuscate the user's exact contextual data.

In this paper, we focus on the \Paco API as it relates to on-loading spatiotemporal contextual storage and subsequent on-device application queries; these elements are depicted within the dashed rectangle in Fig.~\ref{fig:system-architecture}. We define the \Paco data structure, based on the existing {\sf R-Tree} spatiotemporal index. We create the {\sf Smart Insert} method and the query interface that defines the \Paco {\sf API}, and we show how {\sf Access Profiles} can be used to restrict the release of potentially private contextual data. We do not implement communication mechanisms to enable D2D sharing and cloud offloading; we could easily connect \Paco to existing mechanisms in these spaces~\cite{kalbarczyk16:xd}. We do not include the {\sf Context Index} in this investigation; for the remainder of this paper, we treat the structure as storing just the space and time of the observations, what we assume to be the ``heavy-lifting''  and most general purpose pieces of contextual indexing. Since our spatiotemporal index is relational, it supports additional context indexing through joins on selection queries. Including the contextual indices as well as the associated contextual ontology is left for future work, and is also addressed in much existing work on context ontologies~\cite{bettini10:survey, wang2004ontology}.
\section{Related Work and Background}\label{sec:relatedwork}
The ability to store, query, and reason about spatiotemporal information is increasingly essential to mobile applications~\cite{biagioni11:easytracker,blanke14:capturing,yan12:fast}. Increasing concerns about location privacy~\cite{sadeh09:understanding} have motivated researchers to explore privacy primitives associated with location sharing~\cite{guha12:koi, toch10:locaccino}, with a focus on determining what location information to share and how to share it in a way that shields the user.  Furthermore, on-loading, or moving storage and processing of data onto the device, has become popular for improving user experience in mobile computing environments~\cite{han13:case, vallina12:when}.  The contributions in this paper couple the idea of on-loading with the indexing of spatiotemporal data to support valuable contextual data storage of any type.  In order to on-load large amounts of context, a proper adaptation of existing spatiotemporal indexing methods is necessary.

Complementary work in mobile context-awareness enables context processing on-device. ACE~\cite{nath12:ace} is a context reasoning framework that executes on top of sensors and the data streams those sensors generate.  ACE attempts to reduce the overhead of context sensing by making inferences about what context streams to collect using application-level information across multiple applications. ACE's inference mechanisms only work for boolean context types (e.g., ``is the user running'') and not continuous ones like a trace of the user's spatiotemporal movement. Further, ACE assumes that the applications access only instantaneous context information and does not support historical introspection across stored context information. In contrast, a data store like \Paco could be layered {\em underneath} an inference engine like ACE providing a richer store of context information on which to make application decisions.

SeeMon~\cite{kang08:seemon} was an early software enabler of on-device continuous context sensing. Similar approaches to adjusting sensing tasks based on application needs and a device's current situation have resulted in a wealth of similar approaches for continuous context sensing~\cite{kansal13:latency, rachuri12:energy, zhuang10:improving}. As an exemplar, SeeMon intelligently adjusts {\em how} context is acquired, specifically by monitoring context {\em changes} instead of continuously sampling and delivering raw values.  Similarly, work related to efficient trajectory sensing dictates how and when to activate various on-board sensors to accurately determine the device's location while incurring low energy overheads~\cite{kjaergaard11:energy}. CoMon~\cite{lee12:comon} extends these ideas beyond a single device and creates a framework by which co-located devices can collaboratively sense the ambient environment, selecting the best sensors to task {\em across} a network of nearby devices. Like ACE, these approaches for intelligent and efficient continuous sensing deliver the instantaneous context; in addition, they focus on selecting the ``best'' set of sensors and sensing parameters to do that. These approaches do not create an on-device contextual {\em history}. As with ACE, \Paco directly complements these efforts; as (spatiotemporal) context information is continuously generated by these approaches, it can then be stored efficiently and queried in the future using \Paco's API.

In contrast to the wealth of approaches enabling efficient context {\em sensing} on commodity smartphones, existing approaches for {\em storing} that information focus exclusively on off-loading the data to the cloud and enabling efficient and expressive access to it there. That is, existing approaches either assume the sensed context information is shipped off of the device to be stored elsewhere or they assume that the device's applications simply discard any context information that is not immediately consumed. This storage gap is exactly \Paco's target. We {\em on-load} the storage of acquired context information, making it available to the device's applications for {\em future} queries while reducing the energy and communication costs of offloading and giving the user direct control over the potentially personal and private spatiotemporal information.

As a starting point for exploring efficient and effective {\em on-device} storage of spatiotemporal data, it is useful to examine the approaches used in server-based systems. Storing spatial data has a rich history in image processing, geographic information systems~(GIS), and robotics. Grid-based approaches, which divide space into regions and insert data into the grid square representative of the data's location, are the most straightforward~\cite{gaede1998multidimensional}. Clever statistical approaches can optimize queries over this data. This type of approach is not well suited for data that is dynamic or data sets with ``hot spots,'' i.e., spatial areas with a high number of data points. In both cases, selecting an optimal grid size is difficult, and grid bounds may not be known {\it a priori}.

The widely used R-tree~\cite{guttman1984r} maintains a balanced structure by representing objects within minimum bounding rectangles and then creating a hierarchical (tree) representation that relates the rectangles to one another. In an R-tree, a rectangle at a node in the tree completely contains any rectangles of any of the node's descendents in the tree; the tree is structured to support queries through simple computations of intersection and containment. An R-tree is especially useful for storing objects encompassing some area, which maps naturally to the abstraction of minimum bounding rectangles.  The most complexity in using an R-tree involves developing algorithms to minimize or eliminate bounding rectangle overlap~\cite{beckmann1990r}, thus avoiding worst-case query performance. Existing R-tree optimizations reduce this overlap by reinserting points~\cite{beckmann1990r} or duplicating objects~\cite{gaede1998multidimensional}. Because the R-tree is designed for on-disk storage, we use it as the foundation of \Paco's on-device spatiotemporal database, which must be persisted on the device to be shared among applications and to extend its lifetime through application restarts and power cycling of the device\footnote{We use the R*tree, which encompasses the discussed dynamic maintenance optimizations}.

While it is quite beneficial for on-disk storage, the R-tree is not as efficient when used for in-memory computations. The $k$-d tree~\cite{bentley1975multidimensional}, on the other hand, is very well-suited for in-memory computations, especially when all of the data items are known ahead of time, and algorithms can be employed to directly construct a balanced tree. A $k$-d tree can be thought of as a $k$-ary search tree that recursively partitions data points along $k$ coordinate axes alternating between the $k$ dimensions.  \Paco leverages these existing storage structures by employing an R-tree as the base data structure stored on disk, complemented by a $k$-d tree of three dimensions (i.e., $k=3$) as a utility structure to support sub-queries computed in memory.

We are not the first to consider storing temporal information alongside spatial information; this is a natural extension for both the $k$-d tree and R-tree.  More generally, combining spatial and temporal information is particularly prevalent in moving object databases~\cite{erwig1999spatio}. Many optimizations have been explored for improving queries across moving objects such as with the binary string prefix-matching used in~\cite{ganti2016mp}.  Temporal aspects may capture a data item's relevance to the state of the database (termed {\it transaction time}) or relevance to the real physical world (termed {\it valid time})~\cite{tansel08:temporal}. We focus on the latter. In contrast to existing work in more general purpose moving object databases, PACO focuses on a data store whose data points all represent just a single moving object: the user who owns the mobile device collects the data.

In support of efficient storage and retrieval of spatiotemporal data for mobile devices, methods can be loosely categorized into those that ``index past positions,'' ``index current positions,'' and ``index current and future positions''~\cite{mokbel03:spatio}. To date, these approaches build almost exclusively on R-trees, and they are thus complementary to our work. However, existing approaches have developed centralized, heavyweight indexes that are not well suited to the lightweight, flexible, and personalizable implementations demanded by mobile devices. Again, this need for a lightweight index that can reside entirely on a user's personal mobile device is \Paco's goal.

\Paco aims to enable efficient real-time querying of spatiotemporal information by applications on the user's device. Therefore, in designing \Paco, it is essential that the data store does not grow too large both because of the limitations of the device and the need to support quick response to applications' queries for spatiotemporal data. To support on-line location-awareness, researchers have investigated approximate query processing~\cite{sun04:querying}, enabling lossiness in data storage~\cite{cao06:spatio, cudremauroux10:trajstore} by storing line segments comprising a trajectory. In contrast, CDR~\cite{lange08:online} uses an on-line trajectory reduction that relies on the moving objects themselves (e.g. mobile embedded sensors) to temporarily store data.  These approaches are related to some methods that we use in \Paco to intentionally discard redundant data items, however, these approaches continue to rely on a central server or other devices capable of handling significant off-loaded storage and computation.  Further, these approaches do not associate the application context with the spatiotemporal data, thus they lack the ability to use this data to respond to complex application-level queries.

\section{\Paco and the \Paco API}\label{sec:structure}
\begin{figure}[!hbt]
	\centering
	\includegraphics[width=\columnwidth]{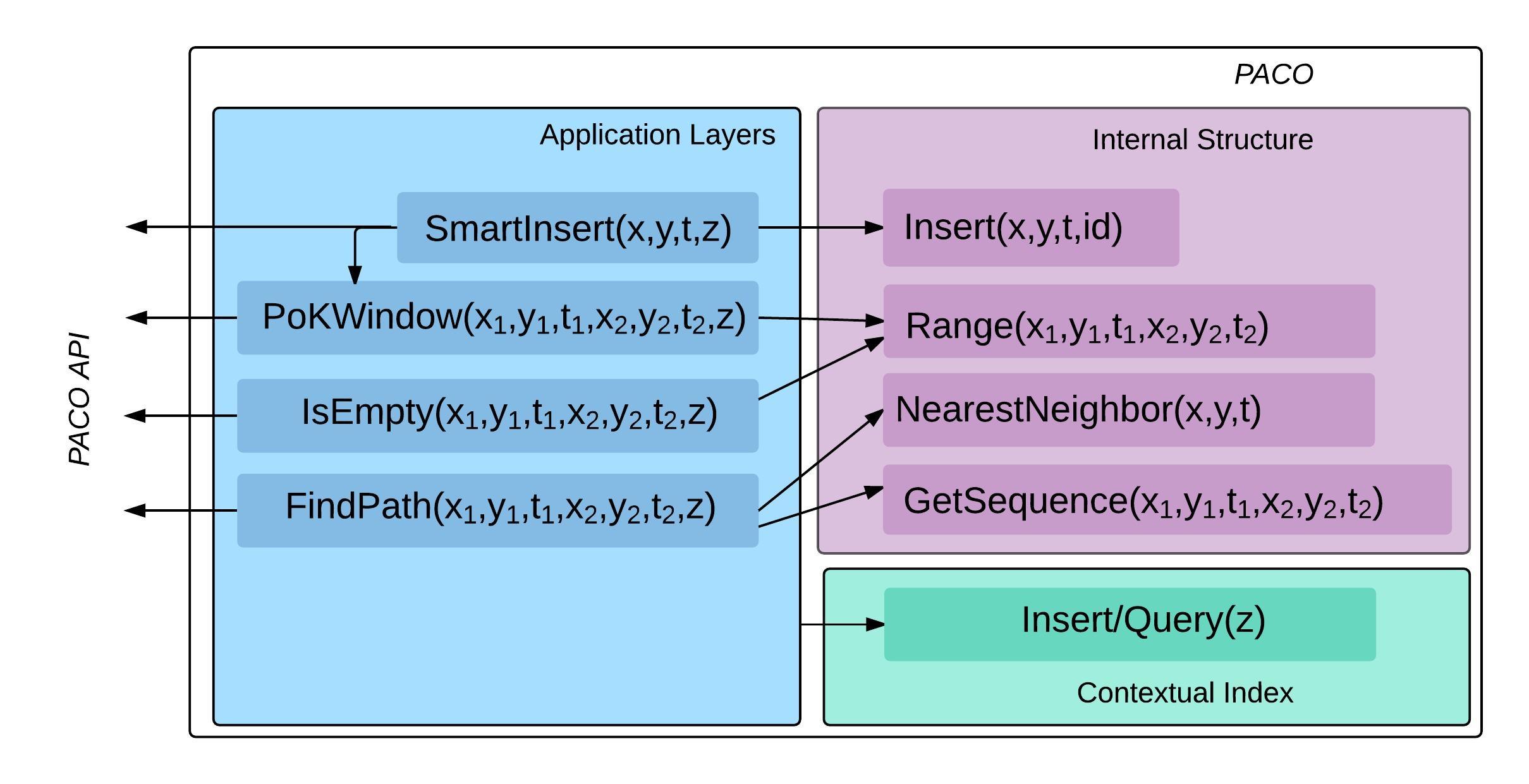}
	\caption{\small Application layers and internal structure of the \Paco API}
	\label{fig:api}
\end{figure}

As shown in Fig.~\ref{fig:system-architecture}, \Paco relies on an internal structure to store spatiotemporal data; because the \Paco API shields the application developer from the specifics of this structure, any spatiotemporal data structure can be used that conforms to the appropriate interface. We first overview the requirements of such a data structure and discuss our use of the R-tree to fulfill these requirements. As shown in Fig.~\ref{fig:api}, the internal structure provides basic query operations across time and space. Upon this structure, we build the \Paco API as a suite of application layers that effectively extend the internal structure's interface, resulting in the complete \Paco abstraction for on-loading a device's spatiotemporal information locally and using that local data storage to answer spatiotemporal queries for applications on the device. Recall that \Paco is intended to replace traditional location services and serve as the primary access point for spatiotemporally-indexed contextual data.  Note that in some cases single raw location data points may need to be exposed, such as for mapping applications that display your current location. We omit this from our discussion of the \Paco API in favor of the more interesting methods that \Paco provides.

\subsection{Required Internal Data Structure Interface}
The internal structure is responsible for the low-level efficient storage of spatiotemporal data.  It must support inserting spatiotemporal data objects in the form $(x, y, t, {\it id})$, where $x$ and $y$ are spatial coordinates, $t$ is a temporal value, and ${\it id}$ is an identifier for the data object that allows \Paco to directly associate the entries in the spatiotemporal data store with the appropriate application-level context entries.  We only use two spatial dimensions for our initial study, but \Paco can be easily extended to support more dimensions. For the remainder of this paper we omit discussion of ${\it id}$ and instead focus on the space and time portions of the indexing; we expect that these will be integral to most future mobile applications and are reasonable targets for storage and query optimizations.

The structure must also support the ability return a list of points matching a range of spatiotemporal values; for example a query may ask for all data points with an area comprising a popular running route from 5-8pm.  This is a common feature of data structures supporting spatial and temporal indexing and is essential to the effectiveness of \Paco's abstractions; as such, optimizing efficiency of this query is crucial. 

In addition, the underlying data structure should support an additional two query operations that will be used in \Paco to support expressive application queries. One is the nearest neighbor operation, which takes a piece of candidate spatiotemporal data and finds the closest matching data item stored in the structure. For instance, applications may ask for the stored data item nearest an historical landmark, closest to given time, or closest to a particular place {\em and} time. Finally, the internal structure must allow for sequential retrieval of time-ordered data items between two query points. This query is of the form $Q\{p_{1},p_{2}\}$, where $p_{1}$ and $p_{2}$ are spatiotemporal points previously inserted into the structure, and the query returns the sequential list of data between these two points. An example sequence query may ask for a list of points between a runner's starting and ending location.

\subsection{The \Paco API}
Building on the internal data structure, we introduce the \Paco API, which establishes a programming interface for answering important spatial and temporal questions and for controlling the contents of the spatiotemporal data store (see the outermost layer in Fig.~\ref{fig:api}). This API is tunable through configuration parameters that can be adjusted to match the current available resources, for different users, and for a particular user's situation or changing environment. While additional operations may be added to \Paco in the future, we demonstrate in Section~\ref{sec:casestudy} the effectiveness of those provided in this work to address existing and future application uses. A basic building block for the \Paco API is \Paco's notion of {\em probability of knowledge}, which we discuss first.

\subsubsection{Probability of Knowledge (PoK)}
Fundamental to \Paco is the creation of a model to represent the amount of knowledge stored in the internal structure about a particular place and time.  Points in the \Paco data store have the form $(x,y,t,z)$ where $x$ and $y$ are spatial dimensions (typically longitude and latitude, respectively), $t$ is the timestamp, and $z$ is an abstraction for any additional context.  In representing additional context as a single dimension $z$ is a significant oversimplification, our structure fully supports expressive extensions including incorporating a full context ontology~\cite{wang2004ontology}. We omit discussion of these details as well as computations on $z$ because the process is more of an exercise in developing the semantics of the context ontology and not fundamental to spatiotemporal data storage and querying. In contrast, the aim of this paper is to perform the ``heavy lifting'' related to the most common context aspects, i.e., space and time.

The mobile applications that motivate our work often want to ask questions of the form: ``How well does the data stored in the structure relate to a given reference point?'' or ``What does the structure knows about a point in space or time?''  For example, a contextual chat application might ask whether a user's historical context indicates purchasing items at a specific store at 5pm in order to connect messages with similarly profiled users.  For a matching data point, $x$ and $y$ may be the GPS coordinates of the store, $t$ the timestamp for 5pm on the current day, and $z$ additional context that indicates shopping (e.g., a recorded purchase). Given some reference point $r=(r_x, r_y, r_t, r_z)$, we might ask what influence a particular stored data point, $p$, has on $r$.  In other words, we want to determine the probability that a point in the structure shares common or related information with the reference point based on their regions of influence. We define this influence on a per-dimension basis, using a spatial influence function $I_s(\Delta x, \Delta y)$ and a temporal influence function, $I_t(\Delta t)$ where $\Delta x, \Delta y$ are the spatial distances from $p$ and $\Delta t$ is the temporal distance from $p$. We say that $p$ {\em influences} any reference point $r$ within this region.  

\begin{wrapfigure}{r}{1.65in}
\vspace{-.5cm}
\includegraphics[width=.45\columnwidth]{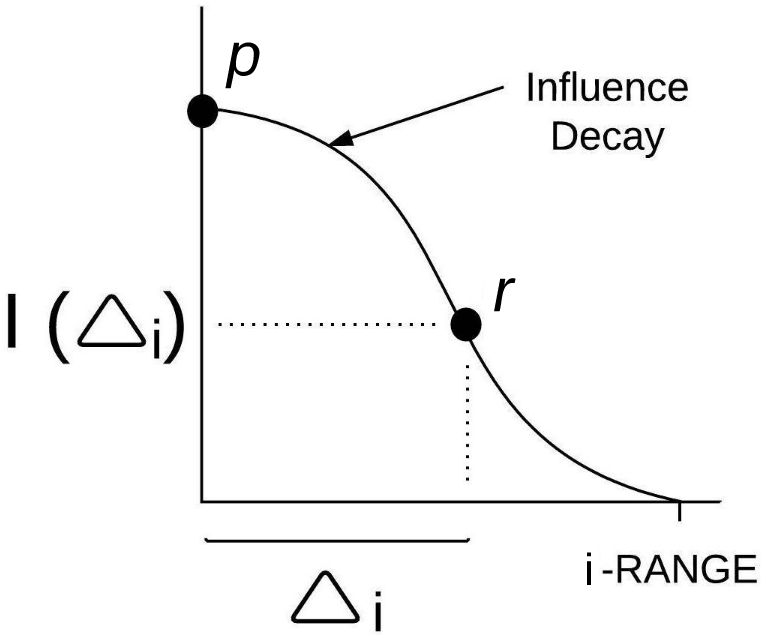}
\caption{\small Influence function for a point $p$ along a particular dimension (i indicates one of the three dimensions)}\label{fig:decay}
\vspace{-.5cm}
\end{wrapfigure} Fig.~\ref{fig:decay} depicts an influence function for some dimension $i$.  Note that the this influence function is similar to modeling a probability density function over space and time, similar to the behavior modeling approach used in \cite{kayacik2014data}.

Using these influence functions, we define the {\em probability of knowledge} (PoK) as the amount of influence a point $p$ has on a reference point $r$ as:
\begin{multline}
{\Delta x = |p_x - r_x|};\:\: {\Delta y = |p_y - r_y|};\:\: {\Delta t = |p_t - r_t|} \\
{{\rm PoK}(p,r) = (I_s(\Delta x, \Delta y) * W_s)  *  (I_t(\Delta t) * W_t)}
\end{multline}
where $W_s$ and $W_t$ are two parameters termed {\sc SpaceWeight} and {\sc TimeWeight}, respectively.  Applications can tune these weights to enable PoK computations to have more or less relative emphasis on spatial or temporal dimensions.  In the remainder of this paper, we fix both values at 1.  

In our study in Section~\ref{sec:benchmarking}, we use linear decay functions. Specifically, we fix the amplitude of the influence functions at 1 and use the value for which the influence function is 0 (x-intercept) or i-{\sc Range} (where ``i'' is either {\sc Space} or {\sc Time}) to adjust the slope of the influence function. The {\sc Range} values can be conceptually thought of as the physical distance and length of time over which a data point should exert influence on reference points.  Note that by using linear influence functions, the decay of PoK behaves quadratically (since the two influences are multiplied in the resulting PoK calculation).

\subsubsection{PoK Window}
\begin{wrapfigure}{r}{1.8in}
\includegraphics[width=.5\columnwidth]{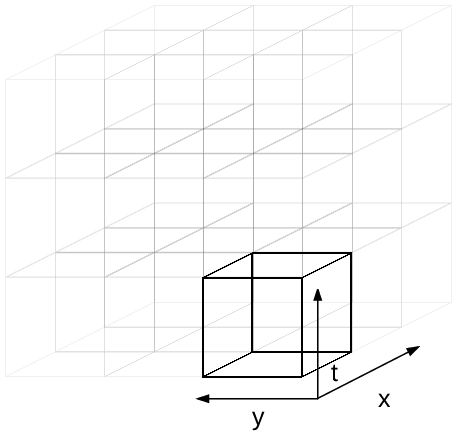}
\caption{\small A PoK window computation highlighting one sub-cube}\label{fig:grid}
\end{wrapfigure}

Potentially more useful than PoK information about a single point is information about the data structure's aggregate knowledge over a spatiotemporal area (i.e., a {\em PoK window}). For example, rather than querying for purchases made at the store at 5pm, an application may query for purchasing knowledge about a reasonably sized area around the store any time in the afternoon. Given such a spatiotemporal region (visualized by a rectangular cube in three dimensions, as shown in Fig.~\ref{fig:grid}), a PoK window query returns the PoK for the {\em region} as opposed to the PoK of a single point; the region's PoK depends on the spatial and temporal influences of the set of ``nearby'' data items stored in the internal structure.

The computation of a PoK window is an integral over the influence functions across the structure's dimensions. In \Paco, we estimate the value by summing the PoK of small {\em sub-cubes} that approximate the influence functions (essentially by performing smaller more tractable window queries), as shown in Fig.~\ref{fig:grid}. The size of the grid sub-cubes trades off computational complexity for the granularity of accuracy of the resulting PoK for the window. To specify the size of the grid sub-cubes, we define a value {\sc GridFactor} that is the number of sub-cubes to include per $i$-{\sc Range} value along any $i$th dimension.  For example, given a {\sc SpaceRange} of 100 meters, a {\sc TimeRange} of 30 minutes, and a {\sc GridFactor} of 2, the sub-cube sizes are $50m \times 50m \times 15min$ (2 cubes per $i$-{\sc Range}).  The value of the {\sc GridFactor} parameter affects both the accuracy and efficiency of the resulting window query; we briefly evaluate these tradeoffs in Section~\ref{sec:benchmarking}.

To compute the PoK of a window, \Paco performs a range query on the internal data structure for each sub-cube. The returned data items are then used to compute the PoK for the sub-cube. Finally, the PoKs for all of the sub-cubes in the window are aggregated to generate the PoK for the window.  Since there are often many sub-cubes for which to calculate PoK values, and, for each sub-cube, \Paco must execute a smaller range query, we perform an initial range query using the entire window's bounds to generate a {\em reference set} of data points for the smaller sub-queries. The initial range query retrieves all of the candidate points stored in the internal structure that are within the window for the PoK computation. We can further restrict the set of candidate points by joining queries that account for filters relative to the additional context information, i.e., $z$. We then execute each sub-cube's smaller window query only on this reference set of candidate points rather than on the entire internal structure (e.g., in our experiments, reported in Section~\ref{sec:benchmarking}, this pre-query eliminated, on average, 97.5\% of the data points from consideration in the larger of our two data sets). \Paco stores this reference set of points in a $k$-d tree that it then uses when calculating each sub-cube's PoK. Our implementation of this approach leverages the benefits of $k$-d trees for in-memory computations and R-trees for maintaining a balanced on-disk database of the entire set of points in the structure.  Our evaluation of PoK window queries in Section~\ref{sec:benchmarking} includes exploring the effectiveness of using a $k$-d tree rather than a simple array of points as the reference structure.

\begin{wrapfigure}{r}{2in}
\includegraphics[width=.60\columnwidth]{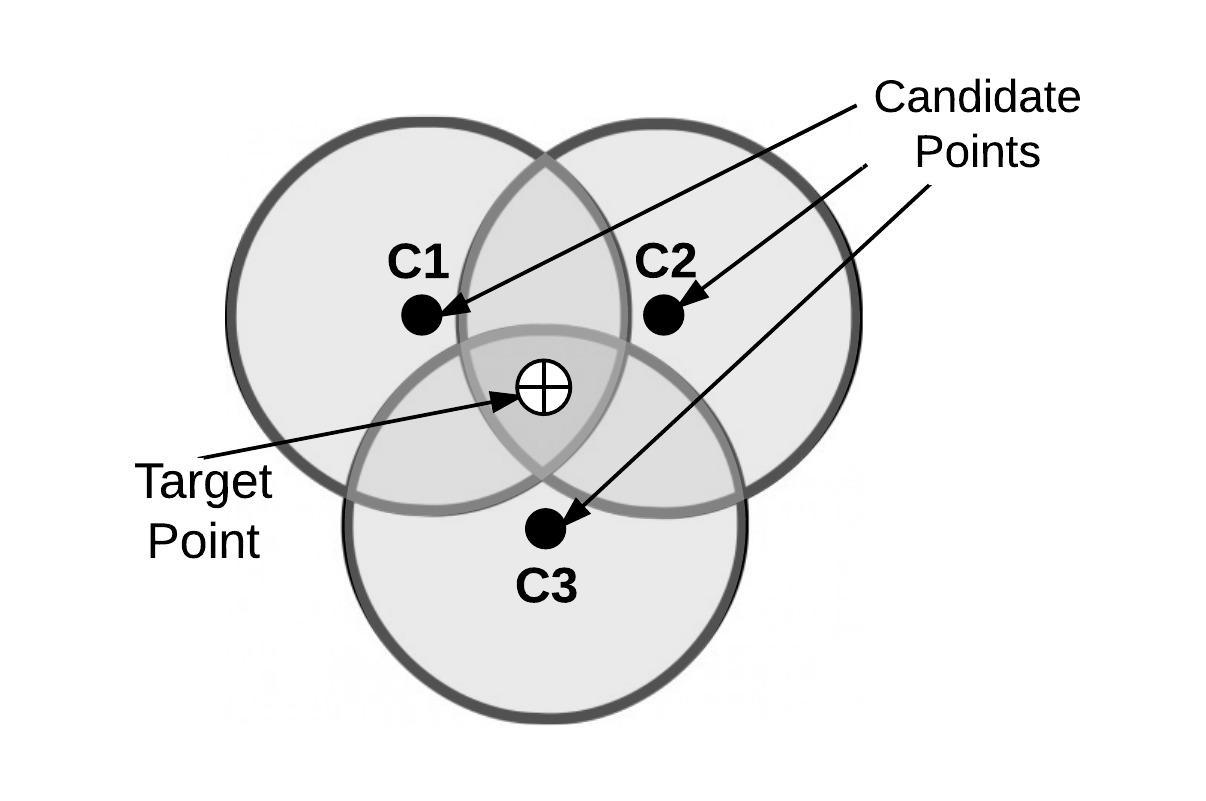}
\caption{\small Inclusion-exclusion and spatial influence}\label{fig:pok2}
\end{wrapfigure}

To calculate each sub-cube's PoK, we need to resolve the influence of potentially multiple data items.  For example, Fig.~\ref{fig:pok2} shows (in only two dimensions) the (overlapping) ranges of influence of three data items C1, C2, and C3.  Computing the PoK for a region that contains all or part of any of the intersection areas in Fig.~\ref{fig:pok2} involves resolving the shared influence between the relevant data points.  

Given a sub-cube, identified by a point at its center (the {\em target point}), we retrieve all of the nearby neighbors of the target point (the {\em candidate points}) using the internal structure's range query. We compute the structure's PoK of the target point using the {\em inclusion-exclusion} principle to account for ``double counting.'' This becomes more complicated as an increasing number of candidate points and dimensions are considered.  In general, the following expression captures the ${\rm PoK}$ at a reference point $r$, where $1\leq i\leq n$ and $K_i$ is the influence from each of the $n$ nearby candidate points. $\mathbf{P}(K_i)$ reflects the distribution for point $i$ after accounting for both the spatial and temporal decays.  Keep in mind that this could be extended to include items from the application-level context (i.e. $z$).

\begin{equation}
{\rm PoK}(r) = \mathbf{P} \left( \bigcup_{i=1}^{n} K_i \right)\label{eq:pok}
\end{equation}

\begin{equation}
\begin{aligned}
{\mathbf{P} \left( \bigcup_{i=1}^{n} i \right)} & =
{\sum_{i=1}^{n} \mathbf{PoK} \left( i \right)} -
{\sum_{i < j}^{} \mathbf{PoK} \left( i \cap j \right)} \\ & \hspace{.5cm} +
{\sum_{i < j < k}^{} \mathbf{PoK} \left( i \cap j \cap k \right)} \\ & \hspace{.5cm} -
{\dotsc} +
{ \left( - 1 \right) ^{n-1} \mathbf{P} \left( \bigcap_{i=1}^{n} i \right)}
\end{aligned}\label{eq:pokdetailed}
\end{equation}
${\rm PoK}$ values are between 0 and 1, inclusive, and the set intersection operator in Equation~\ref{eq:pokdetailed} generates the combined probability for the considered candidate points.  Using Equations~\ref{eq:pok} and~\ref{eq:pokdetailed}, \Paco computes the ${\rm PoK}$ for each sub-cube. We then compute $T_o$ as the sum of the ${\rm PoK}$ values for all of the sub-cubes and $T_P$ as the maximum possible ${\rm PoK}$ (equivalent to every sub-cube having a ${\rm PoK}$ of 1, or, simply, the number of sub-cubes). We return the PoK for the window as the fraction $T_o/T_p$.  For realistic data, our evaluation has show that values of $> 60\%$ indicate high levels of coverage.

Note that Equation~\ref{eq:pok} effectively generates all combinations of the $n$ PoK values.  To mitigate state explosion, we sort the candidate points and select the first {\sc TrimThresh} number of points with the most significance and use these points to compute the window's PoK. In our evaluation, we fixed {\sc TrimThresh} at 10 because it empirically allows for a good number of points to be considered without generating too many sets of combinations.

PoK window queries are flexible and, like their underlying range queries, can be performed over any combination of the dimensions. For example, a query might ask for the PoK over a spatial-only region such as on a college campus or a temporal-only region such as from 9am to 5pm on a particular day. This distinction defines only the bounds for the initial query on the internal structure for which the reference structure is constructed.  The PoK calculation for each sub-cube is unaffected as it still computes the influence values from all three dimensions. Similarly, the PoK computations could be extended to account for dimensions of the application-level context (i.e., $z$).

\subsubsection{Smart Insert}
Maintaining a data structure that contains detailed historical space-time indexes of contextual data generates large structures that may be expensive to maintain and query.  Many data items are redundant in space, time, or both, and maintaining such redundant data items does not add much additional contextual information of use to applications. Most users stay in a relatively confined location throughout a day at work, so updating space-time information on a fixed (e.g., one second) schedule is often overkill.  On the other hand, many mobile location services only update observations if the mobile object has moved a significant distance. This does not account for the temporal relevance of data objects, nor does it account for motion paths where an actively mobile user is constantly looping back on already ``covered'' areas.

\begin{figure}[!hbt]
	\includegraphics[width=\columnwidth]{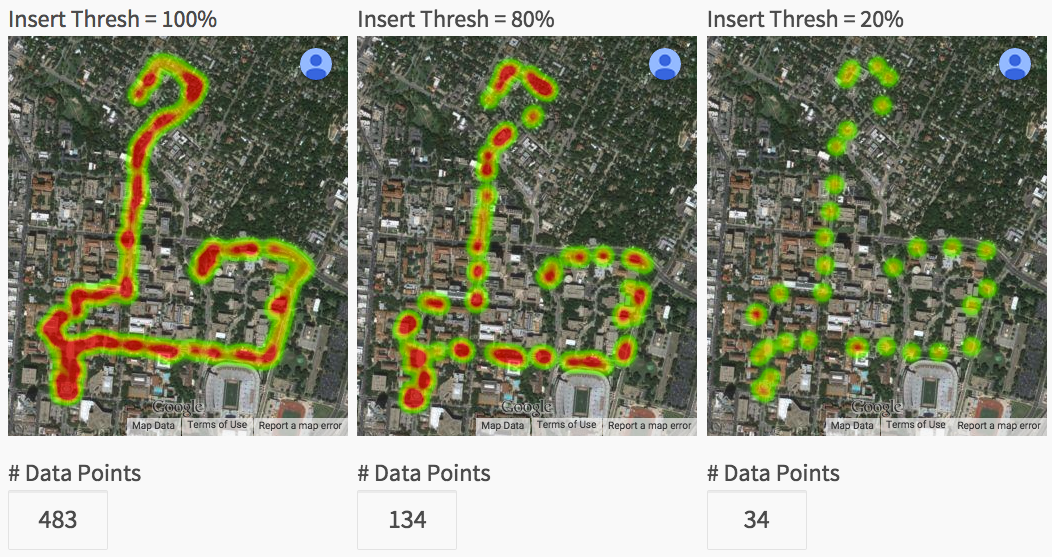}
	\caption{\small Effect of smart insert on spatiotemporal trace data.}
	\label{fig:visual1}
\end{figure}

In \Paco, we introduce {\em smart insert}, which uses application tunable guidance to ask ``how well is this information already represented?'' before inserting a data item. Smart insert accomplishes this check by performing a small PoK window query around the candidate point.  The result of this window query sample is compared to an application-defined parameter, {\sc InsThresh}.  Since both the value of the window query and the value of {\sc InsThresh} are PoK's, they represent the combined spatial {\em and} temporal relevance of the structure relative to the point that is to be inserted.  For example, an {\sc InsThresh} value of 80 means that, if ${\rm PoK(p)} \geq 0.8$ for some new point $p$, then $p$ should not be inserted. Lower thresholds result in a greater reduction in the number of data items inserted, trading some degree of accuracy for storage and computational efficiency as depicted in 
Fig~\ref{fig:visual1}, which gives heat maps of the PoK values of the resulting structure, given the ground truth in the left most figure (i.e., all data samples are inserted, regardless of their contribution to knowledge) vs. smart insert using an {\sc InsThresh} value of 80\% (in the center) and an {\sc InsThresh} value of 20\% (on the right). We provide a more controlled and detailed study of the impacts of these parameters on smart insert and the resulting available knowledge in the data structure in Section~\ref{sec:benchmarking}.

Since this operation uses the PoK window operation described above, its performance is partially dependent on the application parameters defined for coverage window. Using the PoK computations defined above, this considers only space and time information when determining whether a data item is redundant; again, PoK computations can be extended to include additional dimensions of context information, also increasing the expressiveness of the determination of redundancy for smart insert. 

\subsubsection{Find Path}
\Paco also enables queries about a mobile user's {\em trajectories} through space and time.  \Paco's find path operation takes two points $(x_1,y_1,t_1),(x_2,y_2,t_2)$. Under the hood, the operation first finds the nearest neighbor of each point in \Paco's internal data structure, then uses the internal structure's {\sf GetSequence} method to return the path between these two nearest neighbor points. To illustrate, first consider a running/exercise application.  A simple example of find path is to query for two known data points, such as the runner's home and a familiar trail he ran two days ago, for which find path can return the sequential list of data points.  A more interesting example of find path is to query for two non-exact data points.  For example, a user who cannot recall his exact running path could query around a trail head and a nearby lake three weeks ago.  While the user may have not passed through those exact points or at that exact time, find path will determine the nearest two points and return the sequential list of items best matching those points.We include the Find Path operation in \Paco to complement the other operations in motivating the use cases. Find Path is not dependent on configuration parameters, and as such, we omit an evaluation in favor of more interesting methods within \Paco.

A challenge for many of the operations in \Paco is setting the parameters to achieve desired application behavior.  \Paco's programming interface is intentionally designed to make it easy for application developers to use defaults for these parameters or to substitute alternative values on a case-by-case basis. A summary of the parameters as well as concrete values we considered in the evaluation in the next section are summarized in Table~\ref{tab:params}. ``cabs'' and ``peds'' refer to the two data sets we use, which will be introduced in the beginning of the next section. In Table~\ref{tab:params}, {\bf Device Tunable} refers to whether a particular device will adjust the given parameter at run time (and under what conditions). {\bf App Tunable} refers to the likelihood or frequency with which an individual application will adjust the given parameter (e.g., for a particular query or sequence of queries).

    \begin{table*}
    \small
\caption{\Paco parameter values}\label{tab:params}
    \begin{tabular}{ | p{1.8cm}  | p{5.2cm} | p{3.5cm} | p{4.0cm} | p{1.7cm}|}
    \hline
    {\bf Parameter} & {\bf Description} & {\bf Study Values} & {\bf Device Tunable} & {\bf App Tunable} \\ \hline\hline
    {\sc SpaceRange}& range of a point's spatial influence &50m~(peds); 1000m~(cabs) & No & Infrequently \\ \hline
    {\sc TimeRange} & range of a point's temporal influence &5min~(peds); 60h~(cabs) & No & Infrequently \\ \hline
    {\sc GridFactor} & grid squares per {\sc Range} value & 1/2, 1, 2 & Subject to energy constraints & Likely \\ \hline
    {\sc InsThresh} & value deemed significant PoK about a given area & 0.1 - 1.0 & Subject to energy constraints and device mobility & Infrequently \\ \hline
    \end{tabular}
    \end{table*}

\subsection{\Paco Access Profiles}
Our initial intention is to store spatiotemporal data on device to make it available to applications running on that device. In this sense, the above API is one that is assumed to be ``open'' to any application running on the same device as \Paco. An obvious next use of \Paco is to enable individual users to control whether and how their personal spatiotemporal contextual data is shared with applications and services on the device and with other individuals or services executing off of the device. To enable this control, \Paco defines {\em access profiles} that use the \Paco API directly to constrain and filter the spatiotemporal data (and ultimately its derived PoK information) in making it accessible.

In \Paco, users employ the \Paco API's parameters to control the accessibility of the spatiotemporal data stored within \Paco. To provide examples of this process, Table~\ref{tab:profiles} gives a qualitative comparison of different access profiles that could be employed by users or devices to control release of their spatiotemporal information.  The {\em open} profile allows complete access to the \Paco API and is suitable for trusted applications (i.e., likely those resident on the local device). The {\em guarded} profile outlines a safe middle-ground, for instance only allowing the find path operation for entries more than 24 hours in the past (preventing the user's exact recent movements from being tracked); restricting the grid cube size (effectively blurring the granularity of spatiotemporal information released); and providing a minimum window size (only allowing window query sizes five times larger than the {\sc Range} in any dimension). This guarded profile serves as an example for device-to-device interactions (based on the presumption that users are willing to share with other co-located users more than they are willing to share publicly on the Internet~\cite{jones08:geographic}), the {\em guarded}. At the most conservative end of these examples, the {\em restricted} profile serves well for offloading to a public server or sharing with unknown users by only allowing lossy data representations, ensuring that the data is sufficiently obfuscated to make the user comfortable in releasing it.  One could imagine switching between profiles not just for the three different modalities of applications shown in Fig.~\ref{fig:system-architecture} but also based on social relationships with the users of the connected devices, the time of the data, the location of the exchange, the user's mood, etc.
   \begin{table}[!h]
\begin{center}
\caption{Qualitative Comparison of Access Profiles}\label{tab:profiles}
    \begin{tabular}{ | l  | l | l | l |}
    \hline
    {\bf Profile} & {\sc\bf  GridFactor} & {\bf Min. Window Size} &
    {\bf Find Path} \\ \hline\hline
    {\em open} & 1/2 - 2 & None & Yes \\ \hline
    {\em guarded} & 1/2 - 1 & 5X {\sc Range} & $> 24h$ \\ \hline
    {\em restricted} & 1/2 & 20X {\sc Range} & No \\ \hline
    \end{tabular}
\end{center}
    \end{table}

\subsection{Android \Paco Implementation}
We implemented the \Paco components depicted within the dashed area in Fig.~\ref{fig:system-architecture} on Android\footnote{The source code for this implementation is available at \url{https://github.com/nathanielwendt/LSTAndroid}.}. The current implementation of \Paco is an Android service that serves as a sufficient prototype for evaluation. It is our vision that \Paco be implemented at the system level to be more natively shared across applications. Our prototype implementation is, however, more suitable to evaluation on commodity devices, especially in a way that can be replicated by others. We use a 3D R-tree as the internal data structure to store each $(x,y,t)$ point; for simplicity of evaluation, our current prototype does not associate the $z$ values with the spatiotemporal samples, but this is a straightforward extension.  The default SQLite build contained within Android does not include the needed R-tree module, so we compiled our own version of SQLite and accessed it through JNI in our benchmark test runner.  SQLite's R-tree module requires insertion of regions, $(x_{min},x_{max},y_{min},y_{max},t_{min},t_{max})$; for a given $(x,y,t)$ point, we duplicate data so that for each dimension $i$: $i_{min} = i_{max}$. We also adapted the $k$-d tree module from the java machine learning library\footnote{\url{http://java-ml.sourceforge.net/}} for use as our in-memory reference tree.  Lastly, to support storing GPS data, we developed a set of GPS library functions for manipulating spatial distance over latitude and longitude while taking into account curvature of the earth, etc.\footnote{Equations: http://www.movable-type.co.uk/scripts/latlong.html}

\section{Benchmarking}\label{sec:benchmarking}
We performed a series of benchmarks on \Paco to better understand its feasibility and efficiency.  To our knowledge, we are the first to explore on-loading spatiotemporal data at this scale and density; for this reason, we evaluate \Paco across various configurations.  We also compare \Paco's internal structure to a single table database approach (which we call StdTable). For both internal structure options, we compare using the $k$-d tree reference tree for window computes (as discussed in Section~\ref{sec:structure}) versus not using it.  Recall that the $k$-d in-memory reference tree optimizes the otherwise expensive repeated window query that computes over sub-cubes.  All evaluations use mobile trace data from CRAWDAD: (1)~a set of 92 traces of 500-2000 on-foot data points, collected at a university in South Korea~\cite{crawdad:kaist} (``Peds''), which is fairly sparse, and (2)~a set of vehicular traces of taxicabs in San Francisco with 500 taxicabs over 30 days~\cite{crawdad:taxis} (``Cabs''), which is fairly dense with distinct highly populated areas.

We used Moto G 1032 Android Devices (Quad-core 1.2 GHz Cortex-A7, Qualcomm Snapdragon 400 chipset, and 1GB RAM).  Determining execution times in Java can be unreliable due to the JVM's JIT compiler and variations in system behavior. We mitigate these concerns using industry guidelines\footnote{\url{http://www.ibm.com/developerworks/java/library/j-benchmark1/index.html}}. We perform all benchmarks on quiescent devices and run warmup samples on the JIT compiler before measuring execution times. Energy benchmarking uses Qualcomm's Trepn Profiler\footnote{\url{https://developer.qualcomm.com/mobile-development/increase-app-performance/trepn-profiler}}, and power measurements are reported from a recorded baseline established per test.  Despite these measures, execution times and energy levels should not be taken as absolute but rather as reasonable estimates of performance and relative measures across parameter settings.

Our evaluation is framed with the following goals: (1)~maintain fast, responsive query execution for on-device spatiotemporal application queries; (2)~reduce the size of stored contextual data to manageable levels (both for supporting the first goal and for reducing \Paco's memory footprint on resource-constrained devices); and (3)~minimize energy consumption of \Paco's operations. In this section, we take the elements of \Paco's API in turn and evaluate them for these three goals. We also investigate some alternative internal operations for their impact on these goals.

\subsection{Smart Insert}
\begin{figure}[!t]
	\includegraphics[width=\columnwidth]{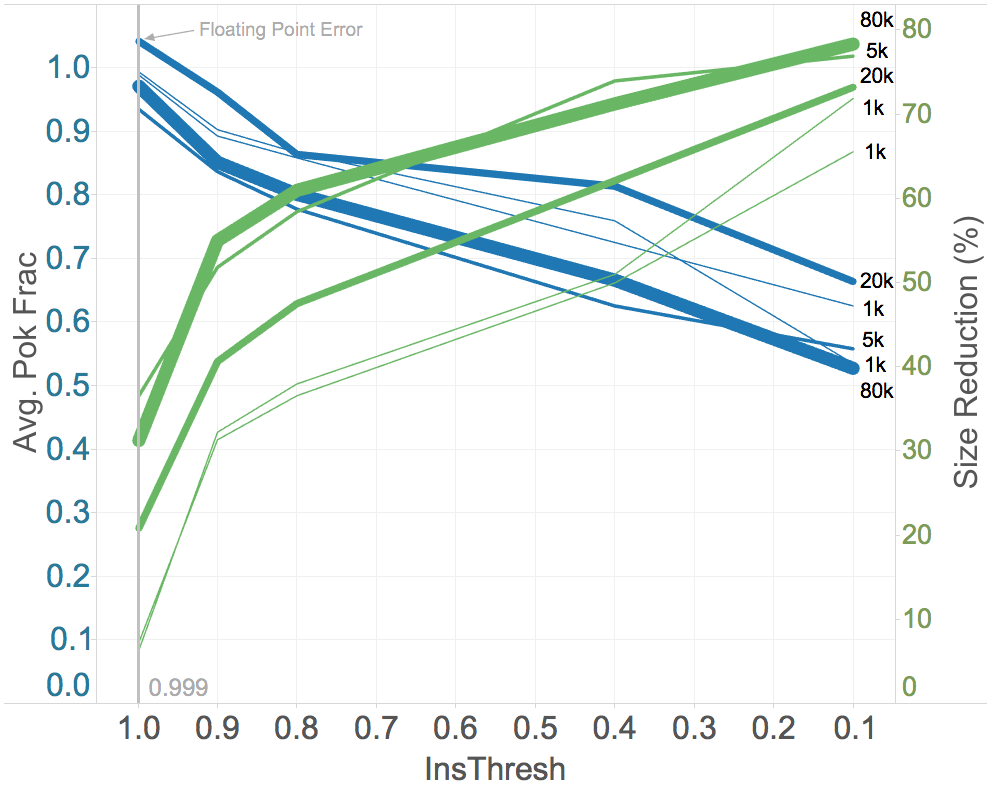}
	\caption{\small Effect of smart insert on PoK and structure size.}
	\label{fig:eval-smart-insert}
\end{figure}

Recall that \Paco's smart insert operation trades accuracy of information for structure size.  With decreasing values of {\sc InsThresh}, \Paco becomes increasingly selective in inserting new data points.  Intuitively, the effect is a decrease in resulting PoK values since there are fewer points to influence region and point queries (i.e., there is less knowledge stored in the structure).  Fig.~\ref{fig:eval-smart-insert} shows the various relationships between {\sc InsThresh}, the size of the internal structure, and the resulting PoK values. A decrease in PoK value indicates a decrease in stored data accuracy from the ground truth of storing all of the points. On this graph, the varying thickness of the lines represents the size of the data set (the thickest line belonging to the largest data set). The left axis shows the average PoK achieved as a function of reducing the {\sc InsThresh}. The right axis shows the degree of size reduction of the overall data structure as a function of reducing the {\sc InsThresh}. As an example, for the largest data set, \Paco achieves an almost 80\% reduction in the size of the data store, albeit at a decreased PoK of about 0.5. 

Larger data sets (such as the trace of 80,000 points) tend to exhibit larger fractional size reductions while still maintaining similar changes in PoK.  We empirically choose a default {\sc InsThresh} value of 0.8 as a reasonable compromise that maintains a high fidelity of the ground truth spatiotemporal information (as measured by the change in PoK) relative to the space savings (which ultimately also result in more efficient querying). At this value, most traces maintain PoK within 80\% of the original value with size reductions of an average of over 48\%.  Obviously, these curves are directly representative of the particular traces evaluated, and as such, some data sets such as 5k do not follow the general trend as they vary in spacing and density of points in the trace. The aim of this portion of the evaluation is to examine real world data sets as a first effort in empirically establishing guidelines for setting configuration parameters.  

\subsection{PoK Window}
\begin{figure}[!t]
	\includegraphics[width=\columnwidth]{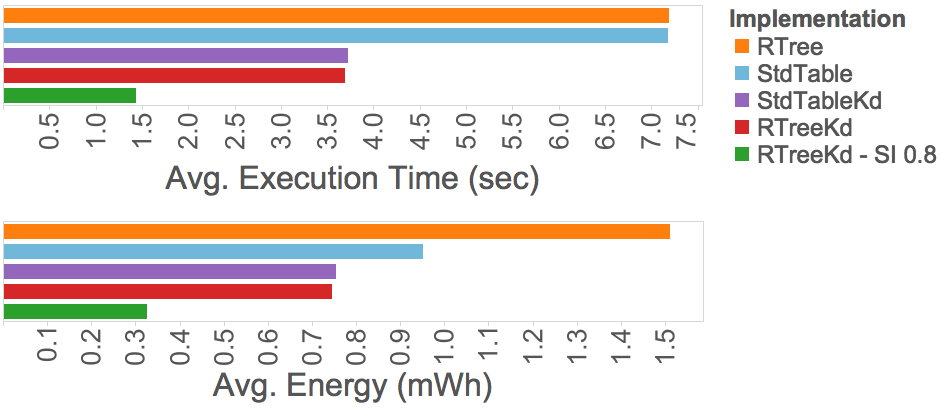}
	\caption{\small Internal structure and reference tree comparison.}
	\label{fig:eval-struct-compare}
\end{figure}

We now turn to examining the PoK window query.  For this evaluation, we used fixed window query sizes of $10{\it km} \times 10{\it km} \times 7 {\it days}$ for cabs and $500m \times 500m \times 10 {\it hours}$ for peds and ran a large number of window queries for each. We choose larger spatial and temporal ranges for the cabs data since queries about vehicular data is likely to cover greater distances and the data stored will be more sparse.  We evaluate using the R-tree and the StdTable (a single table for storing the spatiotemporal data, which represents a conventional baseline) as alternatives for the internal structure and whether or not the $k$-d tree was used for the queries over sub-cubes (utility structure).  We also evaluate the final \Paco option, which uses the R-tree as the internal structure, the $k$-d tree for the sub-cube queries, and a smart insert {\sc InsThresh} parameter of 0.8. We evaluate the average execution time and consumed energy, shown in Fig.~\ref{fig:eval-struct-compare}, across all traces and windows for the 5 possible structure combinations. The raw time values seem high (relative to desired application responsiveness rates) because the evaluation includes very large window sizes that can encompass a large number of data points.  These windows push the limits of the structure; more practical application queries can expect to use smaller windows with fewer than 1000 data points.

The R-tree and StdTable without the $k$-d tree for in memory computation perform the worst (averaging 7 seconds and 1.5 mWh per query). The structures that include the $k$-d reference tree perform significantly better.  The R-tree's benefits over the table approach are not as apparent since the sub-cube computation outweighs the initial range query on the internal structure.  The R-tree's efficiency is more evident in very large  but sparse data sets (i.e., those with many data points but only a small number of those points within a given query window).  The fifth combination uses the R-tree with the $k$-d tree but with a smart insert value of 0.8 (as opposed to 1.0 for the other structures).  As was previously demonstrated, an {\sc InsThresh} value of 0.8 maintains reasonable PoK values while drastically reducing the size of the structure; this is borne out here as well, where reducing the {\sc InsThresh} has a significant impact on both the query time and energy consumption.

\begin{figure}[!hbt]
	\includegraphics[width=\columnwidth]{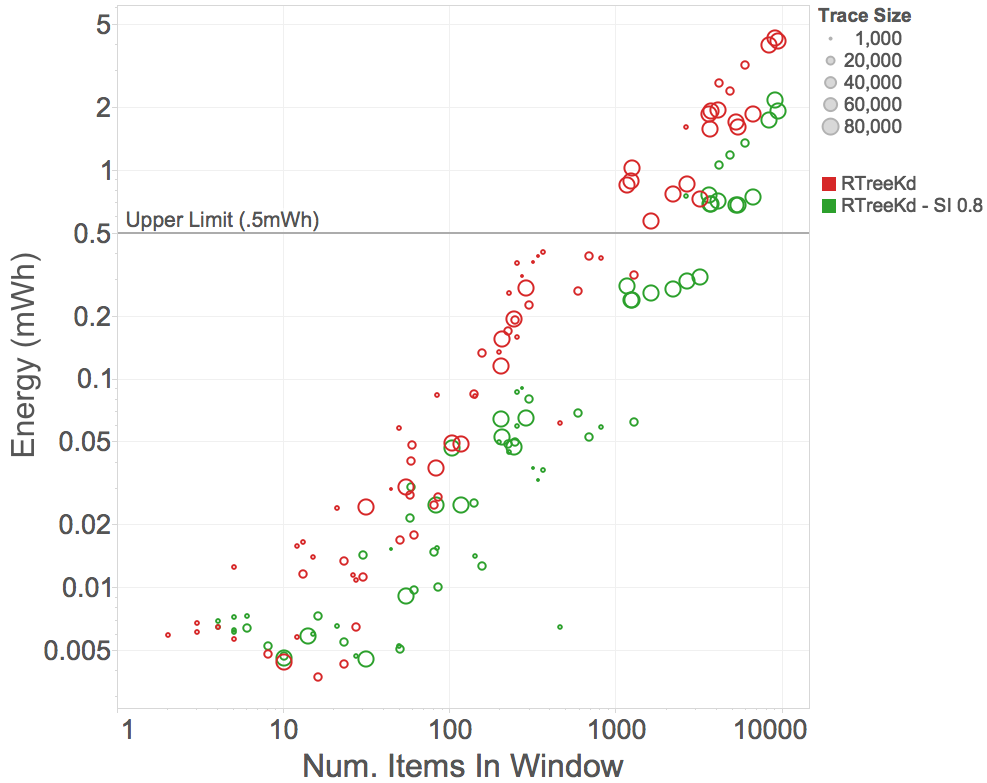}
	\caption{\small Consumed energy vs. window size. Note log axes.}
	\label{fig:eval-supergrid-energy}
\end{figure}

The resources demanded for computing a PoK window query are directly related to the number of data items in the window.  Since the window query is a common operation, it is important that it is effective across a large variation of window sizes. Fig.~\ref{fig:eval-supergrid-energy} shows the energy consumed by a PoK window query for various window sizes.  We show the best structure combination, R-tree with a $k$-d reference tree, both with and without smart insert.  For energy consumption, we set an upper limit of 0.5 mWh as a target for per-query energy consumption, hence the horizontal line in the figure. The Samsung Galaxy S5 has a 2800 mAh battery rated for 3.85 volts.  Estimation yields a total capacity of 10,780 mWhs for the total battery, which allows for 21,560 window queries at this upper limit.  We would like to keep most queries below this upper limit, and as shown in Fig.~\ref{fig:eval-supergrid-energy}, most realistic queries consume about 0.05 mWh of energy, increasing the number of possible queries in a single charge to over 200,000.  Essentially, a user could perform 1,000 PoK window queries per day with less than a 0.5\% effect on battery life.

We similarly evaluated execution time across various window sizes and the data trends very similarly to Fig.~\ref{fig:eval-supergrid-energy}.  We set a realistic target time of 2 seconds since applications should try to keep response times around 1 second~\cite{nielsen1994usability}.  Most windows we evaluated maintained window query execution times below this target, with PoK window queries running for about 0.2 seconds on average.  Also similar to our energy evaluation, the smart insert {\sc InsThresh} value decreased execution times considerably to assist in meeting the 2 second target.

\subsection{Grid Factor}
As important as {\sc InsThresh} is to smart insert, {\sc GridFactor} is to computing the window query.  {\sc GridFactor} determines the number of grid cubes that are used in a window query computation.  Larger number of cubes are desirable from an accuracy standpoint but can add significant computational overhead.

For dense query windows (i.e., those with 500+ data points to consider) we found that a {\sc GridFactor} of 2 averages over 37 seconds in execution time, which is unreasonable for mobile applications.  For windows with fewer points, a {\sc GridFactor} of 2 is more reasonable (resulting in execution times of ~2 seconds).  Ultimately, a {\sc GridFactor} value of 1 is best for default use, with the option to increase the {\sc GridFactor} for small windows or queries where the data is expected to be sparse; one can even consider decreasing the accuracy of the window query result by setting the {\sc GridFactor} to a value of 0.5 when applications require only a lossy representation of a window.

\subsection{Cloud Offload}
We investigate general power consumption of using \Paco in comparison to offloading to a server to demonstrate that our entire approach of on-loading is feasible and reasonably comparable to existing cloud-based silo approaches. In particular, we compare using \Paco~{\em on-device} to entirely offloading all of the device's spatiotemporal data to a cloud, where we assume queries can be processed quickly. Specifically, implementing the same storage features found in \Paco in the cloud in not necessary since we assume that cloud resources can be scaled to serve any desired minimal response time. To mimic minimal server response times, our evaluation sets the service instance to wait for 100ms before returning a response to simulate an optimistically realistic cloud computation. To generate results for the overall energy consumption of insertions and window queries on our cloud instance, we sent a large number of generic HTTP requests and computed the average cost per request.  HTTP requests are sent each time a new location is sampled.

    \begin{table}[!bt]
\caption{\small System storage and power usage }\label{tab:cloud-offload}
    \begin{tabular}{ | p{4.1cm}  | p{1.9cm} | p{1.5cm} |}
    \hline
    {\bf Evaluation } & \Paco~{\bf On-load} & {\bf Cloud} \\ \hline\hline

    {\sc Walkabout Insert (30 min)} & 9.86 mWh & 11.47 mWh \\ \hline
    {\sc Trace Insert (per point)} & 5.87 $\mu$Wh & 7.9 $\mu$Wh \\ \hline
    {\sc PoK Window (per window)} & 25.7 $\mu$Wh & 7.9 $\mu$Wh \\ \hline
    {\sc 20,000 data storage } & 1.24Mb & N/A \\ \hline
     \end{tabular}
\end{table}

To generate the data for this experiment, we performed a one hour ``walkabout'' in which we walked around a WiFi enabled university campus while continuously polling Android location services on the device.  As shown in Table~\ref{tab:cloud-offload}, inserting the spatiotemporal data points into \Paco was about 14\% more efficient than offloading the same data to the cloud, consuming 9.86 mWhs of energy vs. 11.47 mWhs in the cloud case. To further isolate these results from other effects of running the \Paco service while the device was in use, we also compared a controlled insertion of the same data from a captured trace, which resulted in a similar relative result between on-loading and off-loading.  The gist of this result, however, is that in a full 16 hour day of continuous location polling, \Paco would only consume about 316 mWhs of energy to store a complete trajectory of the user's movement.  With modern device battery capacities of over 10,000 mWhs, this is very feasible.

We also compared \Paco's average cost per point insertion and per PoK window query with cloud off-loading of the same activities.  As also shown in Table~\ref{tab:cloud-offload}, \Paco is more energy efficient for data insertion but about three times less efficient for PoK window queries.  This is intuitive since the cloud instance has potentially limitless resources to answer expensive queries but has an overhead associated with each request.  We limited the queries to windows smaller than 1000 data points to better represent realistic application queries. Because it is expected that insertion will be much more common than window queries, the main take away here is that, while \Paco is not designed to save energy, its (overall) energy performance is reasonable in comparison to a conventional approach that relies entirely on a cloud backend for spatiotemporal storage. The last item in Table~\ref{tab:cloud-offload} outlines the storage requirements for 20,000 spatiotemporal data points, for which the \Paco representation required 1.24Mb. Devices could insert a new data point every 5 seconds for a week and only use about 7.5 Mb of storage, which is very reasonable given today's device storages of 16-32Gb.  Note that this leaves plenty of storage for the additional $z$ context data we also plan to include.

\section{Use Cases}\label{sec:casestudy}

\definecolor{dkgreen}{rgb}{0,0.6,0}
\definecolor{gray}{rgb}{0.5,0.5,0.5}
\definecolor{mauve}{rgb}{0.58,0,0.82}

\lstset{frame=tb,
  language=Java,
  aboveskip=3mm,
  belowskip=3mm,
  showstringspaces=false,
  columns=flexible,
  basicstyle={\scriptsize\ttfamily},
  numbers=none,
  numberstyle=\tiny\color{gray},
  keywordstyle=\color{blue},
  commentstyle=\color{dkgreen},
  stringstyle=\color{mauve},
  breaklines=true,
  breakatwhitespace=true
  tabsize=1
}

In this section, we provide two case studies for demonstrating applications' use of the \Paco service. The first of these examples takes an existing application and refactors it to remove its siloed use of spatiotemporal information to replace it with the use of \Paco's abstractions instead. The second shows novel application behavior that is enabled by the existing of \Paco.  In both uses cases, the device employs \Paco's smart insert method to collect spatiotemporal context information, and the device (user) itself manages the insertion thresholds. To provide more application richness, these example applications also bring back the $z$ component of application-level context.

\subsection{Running}
The first example considers a running application that collects location samples for a specified period of time (explicitly started and ended by the user) to collect and report on path information, pacing, and route times. Many such applications exist, including the MapMyRun\footnote{\url{https://www.mapmyrun.com}} family of applications, the Nike+ app\footnote{\url{http://www.nike.com/us/en_us/c/nike-plus/running-app-gps}}, or even apps like Spotify Running\footnote{\url{https://www.spotify.com/running}} that use additional forms of context information (e.g., running tempo) to deliver media content. In all of these applications, context collection is a dedicated part of the app. If a user wants to employ all three of these applications simultaneously (which is reasonable, given that they all provide different capabilities, connections, and statistics), each device maintains its own store of spatiotemporal and context information, and each of these applications pushes that personal context information to a cloud service for processing and storage. Much of the context-related activities of the applications are redundant, as is the cloud offloading, and the latter also releases potentially private information.

In this \Paco use case, we therefore show how one would refactor such an application to make use of \Paco's spatiotemporal context abstractions. To start with, because \Paco continuously collects spatiotemporal context information, the user no longer has to remember to open each of the apps and explicitly select ``begin run.'' Instead, the device passively and unobtrusively records the context information (based on its settings) for later querying by the applications. That is, based on the context collected by \Paco on the device, the application could recreate a run {\em post hoc} based on the information stored in the \Paco structure. Fig.~\ref{lst:runningapp} shows code that would be launched as startup of a running statistics app (e.g., like mapmyrun) to examine route and pacing information for a previously performed run.

\begin{figure}[!hbt]
\begin{lstlisting}
//on app startup
ContextWindow window = 
    ContextWindow.Builder().setTimeWindow(lastAppUse, now)
		                   .setContextMobility(``Running'')
		                   .build();
Paco.setGridFactor(0.5);
double pok = Paco.windowPoK(window);
if(pok > 0.1){
    List<ContextPoint> points = Paco.findPath(window);
    // display map view and/or share running profile
}
\end{lstlisting}
\caption{\small Querying for a activity in a running app.}
\label{lst:runningapp}
\end{figure}

As shown in Fig.~\ref{lst:runningapp}, when the user opens the app, it can access relevant context pieces from \Paco's historical record.  The app first creates a context window with rough precision.  \Paco's configurability allows for a lossy representation of PoK for this query since the exact value is not important, and a general indication above some base threshold (0.1) suffices.  This allows for a fast qualifying query before the app queries for the exact path information.  Once the latter is obtained, the path can be displayed to the user or offloaded to the cloud for use across devices or for archiving.  The user can also allow this application access to specific contextual data.  For instance, the application may be only allowed to offload context that the user has explicitly provisioned, in this case spatiotemporal data with a mobility profile. Instead of offloading the user's entire trajectory, however, the app can offload exactly only the data from the running activity and then only at a level of abstraction matching a stated access profile.

Given that, for example, accelerometer data is part of the $z$ context collected by the device and associated with the spatiotemporal samples, a more sophisticated running statistics app can give the user historical information about running tempo throughout the entire path. Similarly, an executing app like a tempo-sensitive music playlist could use the information from \Paco~{\em as it is collected} to adjust the songs selected as part of the playlist. That is, \Paco supports {\em both} applications' needs for instantaneous context information and the ability to query context information retrospectively.

\subsection{ContextChat}
To demonstrate \Paco's ability to support next-generation applications by enabling new uses of context, we turn to a {\em ContextChat} example, which connects messages between users not only co-located in time and space but across other contextual attributes, for instance based on their exercise history, their dining patterns, or the spending profiles. To demonstrate the types of interactions that such an application might have with \Paco, we consider a use case in which the application aims to send chat messages between drivers who have similar driving experiences. In this example, the selection of similar drivers is based on their spatiotemporally captured paths while engaged in the higher level context activity of ``DRIVING.'' However, using even more expressive notions of context (i.e., $z$), the selection of similar drivers could also be based on other context measures, for instance, of a driver's level of aggressiveness. 

Fig.~\ref{lst:contextchatapp} demonstrates how this application would capture the similarity between the drivers (i.e., as measured by PoK) within given space and time bounds.  For demonstration purposes, this example uses a static query, but more advanced queries could query over dynamic properties determined over user experience.

\begin{figure}[!hbt]
\begin{lstlisting}
SpaceWindow spaceWin = SpaceWindow.bound(currentLoc, 100);
    //100 meter bound

ContextWindow window = ContextWindow.Builder()
                   .setTimeWindow(now - Time.10MINS, now)
                   .setSpaceWindow(spaceWin)
                   .setContextMobility(ContextMobility.Driving);

Paco.setGridFactor(2.0);
double pok = Paco.windowPoK(window);
//retrieve messages
List<Message> messages = 
    CloudBackend.GetMessages(userId, window, pok);

//send a message
CloudBackend.SendMessage(userId, window, pok, message);
\end{lstlisting}
\caption{\small Sending and receiving messages in the ContexChat app.}
\label{lst:contextchatapp}
\end{figure}

First, the app creates a space window for a 100 meter perimeter around the user's current location.  Using this space window, relevant time, and driving context, the app creates a contextual query window. The precision is set to fine since the actual PoK value is important.  The PoK value and the query are sent to a backend chat service that retrieves any messages sent by other drivers that match the given context.  No specific contextual data is sent to the siloed cloud backend, simply a region and a configurably lossy PoK value representing coverage for a contextual query.  This information is abstracted but is sufficient for a backend to index and query to send messages to appropriate users. While this example uses a centralized cloud backend to exchange the chat messages, \Paco is also compatible with D2D uses.  Devices could share context windows, requesting data queries from nearby users directly to gain information about their surroundings, what they are doing, and where they are heading.

\section{Discussion}\label{sec:discussion}
We have outlined how \Paco provides novel types of spatiotemporal queries while intelligently limiting the footprint of the structure on resource constrained mobile devices through the smart insert operation.  While checking the usefulness of the data to \Paco before inserting it is a good first step, extension of \Paco could use a context-aware location sampling technique such as only sampling detailed location when an accelerometer indicates significant movement or by inferring expensive location sampling from cheaper sensors such as in ACE ~\cite{nath12:ace}.  Such a step, performed before even considering whether a sample is a candidate for insertion in the \Paco data store, has the potential to provide energy savings on resource constrained mobile devices. Furthermore, \Paco enables control for the lossiness of the stored structure as well as lossiness in query results.  The lossiness of the query results can be dynamically tuned per-application, but the lossiness of the stored structure must match the greatest accuracy requirement of all the applications.  Going forward, \Paco might support early subscription of accuracy requirements to be able to tune the lossiness of the stored structure.  On the other hand, because \Paco brings all of the spatiotemporal collection under one roof, each spatiotemporal data point is collected only once. This is in contrast to conventional approaches in which multiple applications may be collecting the same spatiotemporal data independently, which is also potentially wasteful.

The work presented in this paper is a necessary first step in giving mobile users control over vast amounts of personal contextual data that is collected about them without unnecessarily limiting the expressiveness of the queries as exposed to applications.  Future work will look to develop a contextual ontology to couple with \Paco's current spatiotemporal indexing, further leverage the internal R-Tree structure to provide faster data region summaries, and to build a system of sharing summaries with other devices or determining appropriate cloud offloading to supplement \Paco.

\section{Conclusions}\label{sec:conclusions}
The key contribution of this paper is to make it possible to entirely {\em on-load} storage and querying of users' potentially highly personal (and private) spatiotemporal context information. To our knowledge, we are the first to enable such on-loading while providing a detailed historical view of spatiotemporal context to applications. Motivated by real-world applications, we introduced \Paco, a programming abstraction that enables efficient on-loading of contextual data storage and querying by indexing crucial spatiotemporal components. We placed \Paco in a broader system context, demonstrating its ability to support D2D exchange as well as cloud offloading while maintaining user control over potentially sensitive data.  We evaluated \Paco over a variety of parameters and compared it to cloud based alternatives, and ultimately demonstrated the efficiency and feasibility of including \Paco on modern devices. Not only does \Paco give users control of their own spatiotemporally tagged context information, but, by making spatiotemporal context storage and querying a {\em system service}, \Paco breaks down the walls between today's existing siloed applications, sharing collected context information among all applications on the device.

 As demonstrated in this paper, our implementation of \Paco enables on-loading contextual storage, giving users direct control over potentially private information and enabling low-latency responses to spatiotemporally indexed contextual queries for a wide variety of applications.

\section*{Acknowledgments}
This material is based upon work supported by the National Science
Foundation under Grant No. CNS-1218232. Any opinions, findings, and
conclusions or recommendations expressed in this material are those of
the author(s) and do not necessarily reflect the views of the National
Science Foundation.

\bibliographystyle{IEEEtran}
\bibliography{paco}

\begin{IEEEbiography}[{\includegraphics[width=1in,height=1.25in,clip,keepaspectratio]{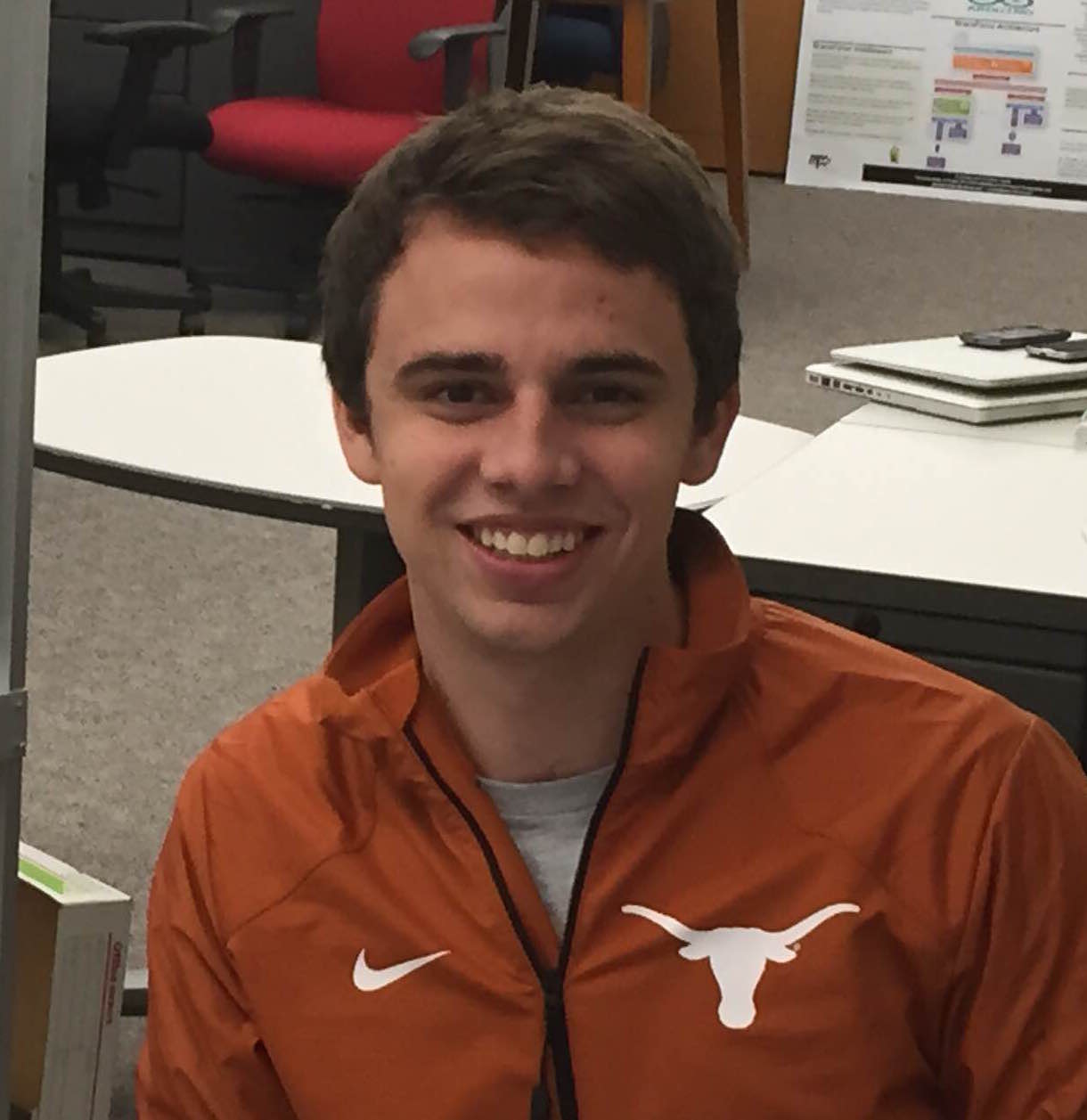}}]{Nathaniel Wendt.}%
Nathaniel is a graduate student at the University of Texas at Austin researching applications of mobile context for preserving privacy, enabling better monitoring and learning in mobile platforms and the IoT.  Nathaniel primarily focuses on middleware that leverages mobile context to enable simplified and more intelligent application programming in pervasive computing environments.
\end{IEEEbiography}

\begin{IEEEbiography}[{\includegraphics[width=1in,height=1.25in,clip,keepaspectratio]{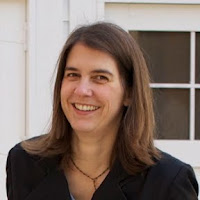}}]{Christine Julien.}%
Dr. Julien is a professor in the Department of Electrical and Computer Engineering at the University of Texas at Austin. Her research focuses on the intersection of software engineering and dynamic, unpredictable networked environments and develops models, abstractions, tools, and middleware whose goals are to ease the software engineering burden associated with building applications for pervasive and mobile computing environments. Dr. Julien's research has been supported by the National Science Foundation (NSF), the Air Force Office of Scientific Research (AFOSR), the Department of Defense, Freescale Semiconductors, Google, and Samsung, and the results have appeared in many peer reviewed journal and conference papers. Dr. Julien graduated with a D.Sc. in 2004 from Washington University in Saint Louis and earned the M.S. degree in 2003 and the B.S. with majors in Computer Science and Biology in 2000 (both also from Wash. U.).
\end{IEEEbiography}

\end{document}